%% file: conference_101719.tex
\documentclass[conference]{IEEEtran}
\IEEEoverridecommandlockouts
\usepackage{tikz}
\usepackage{pgfplots}
\usepackage{subcaption}
\usepackage{graphicx}
\usepackage{lipsum}

\pgfplotsset{compat=1.18}
\usetikzlibrary{arrows.meta, patterns}

\newlength{\figwidth}
\usepackage{cite}
\usepackage{amsmath,amssymb,amsfonts}
\usepackage{algorithmic}
\usepackage{graphicx}
\usepackage{textcomp}
\usepackage{xcolor}

\def\BibTeX{{\rm B\kern-.05em{\sc i\kern-.025em b}\kern-.08em
    T\kern-.1667em\lower.7ex\hbox{E}\kern-.125emX}}
\begin{document}

\title{Energy-Efficient and Actuator-Friendly Control Under Wave Disturbances: Model Reference vs. PID for Thruster Surge}

\author{Anıl Erdinç Türetken, Hakan Ersoy, Aslıhan Kartci \\
\\
Yildiz Technical University \\
Department of Electronics and Communications Engineering \\
Istanbul, Türkiye \\
\\
\text{erdinc.turetken@std.yildiz.edu.tr, hakan.ersoy@yildiz.edu.tr,aslihan.ozakin@yildiz.edu.tr}
}

\maketitle

\begin{abstract}
In this study, we compare a model reference control (MRC) strategy against conventional PID controllers (tuned via metaheuristic algorithms) for surge velocity control of a thruster-driven marine system, under combined wave disturbance and sensor noise. The goal is to evaluate not only tracking performance but also control energy usage and actuator stress. A high-order identified model of a Blue Robotics T200 thruster with a 2~kg vehicle is used, with an 8~N sinusoidal wave disturbance  applied and white noise ( added to the speed measurement. Results show that the optimized MRC (MRC-R*) yields the lowest control energy and smoothest command among all controllers, while maintaining acceptable tracking. The IMC-based design performs closely. In contrast, PID controllers achieve comparable RMS tracking error but at the cost of excessive actuator activity and energy use, making them impractical in such scenarios. Future work will involve experimental validation in a water tank to confirm these findings.
\end{abstract}


\section{Introduction}
Marine vehicles and thruster-driven platforms operating in real seas face significant challenges due to environmental disturbances and sensor noise. Waves, currents, and wind induce time-varying forces that complicate the control task~\cite{1}. In such conditions, maintaining robust performance with a basic PID controller is difficult. Another practical issue is measurement noise: a PID’s derivative action tends to amplify high-frequency noise, leading to erratic “chattering” of the control output. This excessive high-frequency actuation not only degrades performance but can also accelerate actuator wear and increase maintenance costs~\cite{2}. Therefore, there is a strong motivation to seek control approaches that can handle disturbances while ensuring smooth and energy-efficient actuator commands. 
Unlike a standard PID relying solely on feedback, the IMC approach can better handle complex dynamics and disturbances while maintaining performance.
In parallel, energy efficiency and actuator longevity have become important considerations, especially for battery-powered or long-endurance underwater vehicles. 

Recent works have begun to include control energy and smoothness directly in controller design. For instance, reinforcement learning and model predictive control schemes for autonomous underwater vehicles have added penalties on control effort and changes, which successfully reduced energy consumption and sudden control fluctuations. These approaches highlight that penalizing rapid actuator movements can yield smoother operation and prolong actuator life. Despite the availability of advanced methods, PID controllers remain ubiquitous in marine control due to their simplicity and ease of implementation. To improve PID performance, modern heuristics and AI techniques are often used for auto-tuning. Evolutionary algorithms such as Particle Swarm Optimization (PSO), Differential Evolution (DE), and Whale Optimization Algorithm (WOA) have been applied to tune
PID gains in marine robotics and other domains. In a recent study on an underwater robotic
manipulator, WOA-tuned PID outperformed PSO-tuned PID, achieving faster settling and lower tracking error (ITAE) than the alternatives. These results underscore that well-tuned PIDs can be competitive in nominal tracking performance. However, there is a gap in understanding how such
optimized PIDs compare with advanced controllers under realistic combined disturbances (wave forces + sensor noise), especially when evaluating metrics of energy usage and actuator stress.

This paper aims to fill that gap by providing a comprehensive comparative study of a model reference control(without optimization cost) vs. optimally tuned PID controllers in a thruster-driven surge velocity control problem under wave disturbance and measurement noise. In particular, we focus on metrics related to energy efficiency (integral of squared control input) and actuator-friendly control (smoothness quantified by the integral of squared input rate), alongside standard tracking performance indices. By doing so, we shed light on the practical trade-offs between aggressive disturbance rejection and actuator health in marine control systems. The following sections present the system model and disturbances, describe the controllers designed (MRC variants, IMC, and PID tuning), then detail the simulation scenario and performance metrics. Finally, we discuss the results.
\section{System Model and Disturbances}
\subsection{Thruster Dynamics}

A linear transfer function for the BlueRobotics T200 thruster was identified from the publicly available dataset in \cite{1.matlab} using MATLAB’s System Identification Toolbox. Model structure and order were selected iteratively to balance fit quality and parsimony, thereby avoiding over-parameterization.

The resulting best-fit model has a second-order numerator and a fourth-order denominator,
\begin{equation}\label{eq:plant_tf}
T(s) ;=; \frac{330.8s^{2} + 16550,s + 5854}
{s^{4} + 1351s^{3} + 18130,s^{2} + 550{,}400,s + 134700},
\end{equation}

i.e., a 2-zero/4-pole structure. This compact representation captures the dominant electromechanical and hydrodynamic effects observed in the experiments—added mass, viscous and pressure drag, and motor/sensor lags—over the operating bandwidth.
Model fidelity was quantified using MATLAB’s System Identification Toolbox “fit to data” metric  computed from residual analysis. As summarized in Table~\ref{tab111}, the transfer function in Eq.~\eqref{eq:plant_tf} achieves an 85.78\% fit on the estimation dataset and 91.27\% on the validation dataset, indicating good predictive accuracy over the operating range.

\begin{table}[htbp]
\caption{Fit-to-data  for the transfer function in Eq.\eqref{eq:plant_tf}}
\label{tab111}
\centering
\begin{tabular}{l c}
\hline
\textbf{Dataset} & \textbf{Fit (\%)} \\
\hline
Estimation & 85.78 \\
Validation & 91.27 \\
\hline
\end{tabular}
\end{table}

\subsection{Plant Model} The test platform is a surge (longitudinal) velocity control system driven by a BlueRobotics T200 thruster. The dynamic model consists of the thruster’s own dynamics in series with the

vehicle’s surge motion (essentially an integrator from force to velocity, given by $1/(m s)$ with mass $m=2$ kg). The thruster was characterized by a high-order transfer function $G_{m}(s)$ identified from
data. The combined open-loop plant $G(s) = G_{m}(s)\cdot \frac{1}{m s}$ was reduced to a minimal
realization for controller design. For brevity, the full transfer function is not reproduced here;
in summary $G(s)$ is a fourth-order system with a pair of lightly damped poles (from the thruster)cascaded with the integrator representing vehicle inertia. This represents a challenging plant with significant dynamics (including actuator lag and hydrodynamic effects) and an integral action making it type-1 (capable of setpoint tracking with zero steady error under ideal control)
\begin{figure}
    \centering
    \includegraphics[width=0.8\linewidth]{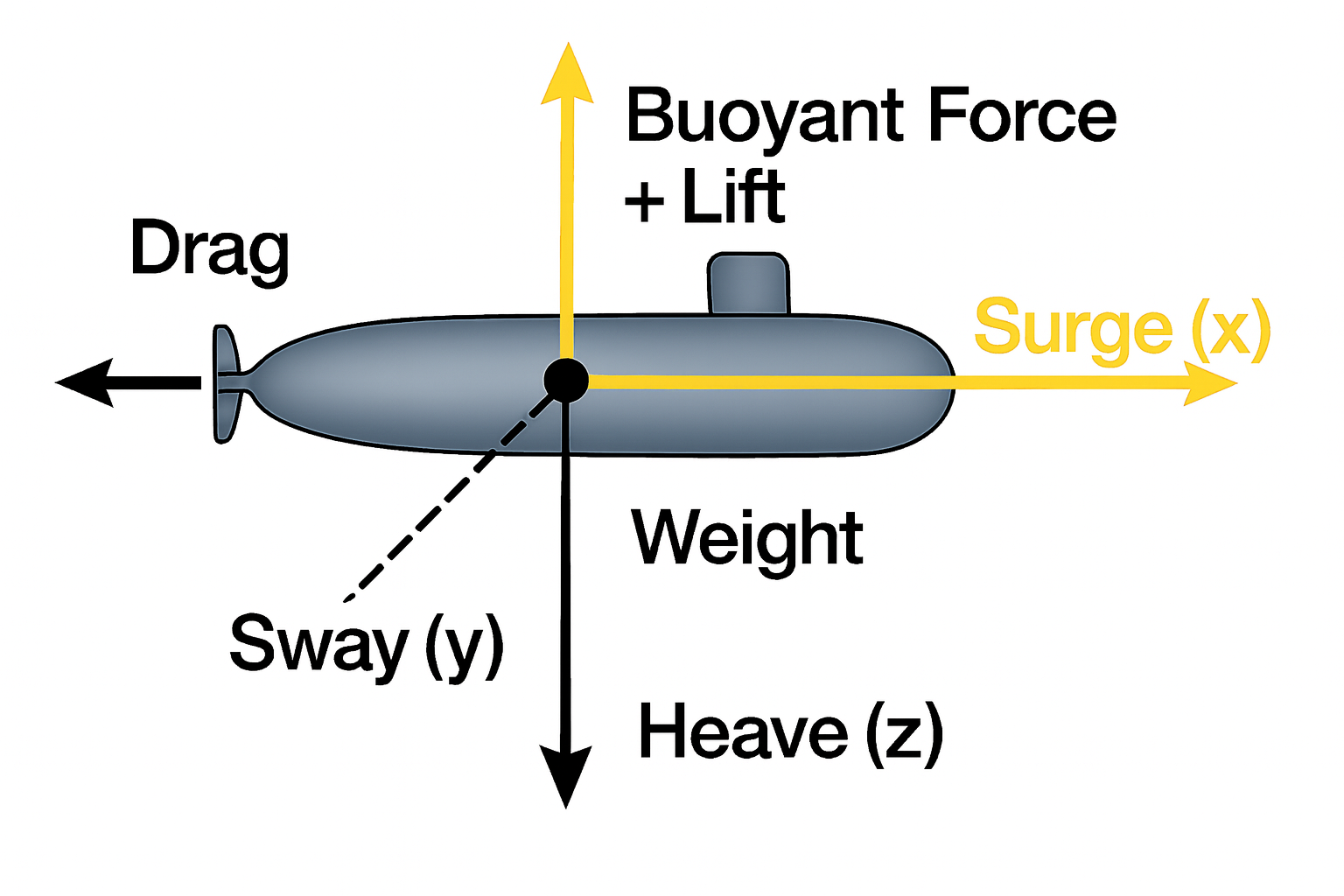}
    \caption{Enter Caption}
    \label{fig:placeholder}
\end{figure}

\section{Surge Motion Control and Plant Model}

In surge motion control the vehicle tracks a commanded forward speed by balancing propulsive thrust against hydrodynamic drag and inertia. A standard one–dimensional balance reads
\begin{equation}
\label{eq:surge_balance}
(m+X_{\dot u})\,\dot u + D_1\,u + D_2\,u|u| \;=\; T(u_c) + F_{\mathrm{dist}}(t),
\end{equation}
where $u$ is the surge velocity, $m$ the dry mass, $X_{\dot u}$ the added mass, $D_1$ and $D_2$ linear and quadratic drag, $T(\cdot)$ the thrust generated by the command $u_c$, and $F_{\mathrm{dist}}$ exogenous disturbances (waves, currents).

\subsection{Closed-Loop Interconnection}
\begin{figure}
    \centering
    \includegraphics[width=1\linewidth]{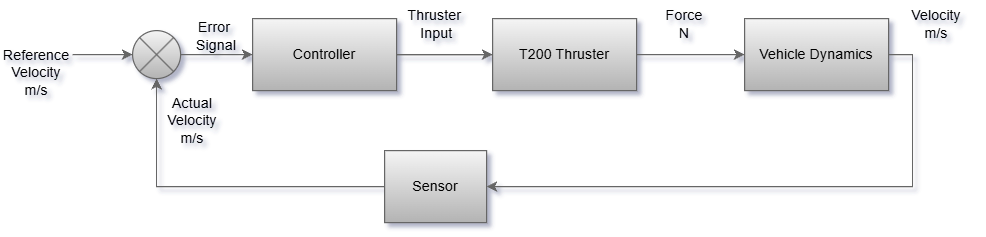}
    \caption{Generic Closed Loop}
    \label{fig:closed_loop}
\end{figure}

The closed-loop architecture in Fig.~\ref{fig:closed_loop} is a generic
unity-feedback speed loop. The block labelled \emph{Controller} denotes the
compensator $C(s)$ (PID/MRC/IMC). The plant comprises the T200 thruster followed
by surge kinematics, i.e., $G(s)=T(s)/(ms)$. Wave disturbance $d(t)$ is injected
additively at the actuator input (not shown for clarity), and measurement noise
$n(t)$ corrupts the sensor reading; thus the controller sees the error in
\eqref{eq:error}.

\begin{equation}
\label{eq:error}
\begin{aligned}
e(t) &= r(t) - y_m(t) \\
     &= r(t) - \big(y(t) + n(t)\big).
\end{aligned}
\end{equation}

with $r$ the reference speed, $y$ the true surge speed and $n$ additive measurement noise. The control output $u$ is injected at the actuator input together with an additive wave disturbance $d(t)$, so that the total input to the plant is $u_{\text{tot}}(t)=u(t)+d(t)$. This convention matches the experimental setup and keeps noise out of the actuation path.

\subsection{Plant Transfer Function}
The thruster dynamics identified from data are represented by
\begin{equation}
\label{eq:T}
T(s)=\frac{330.8\,s^{2}+16{,}550\,s+5{,}854}{s^{4}+135.1\,s^{3}+18{,}130\,s^{2}+550{,}400\,s+134{,}700},
\end{equation}
a two-zero/four-pole model. Surge kinematics add the physical integrator $1/(ms)$ with $m=2~\mathrm{kg}$, yielding the plant used for design and simulation:
\begin{equation}
\label{eq:G}
G(s)\;=\;T(s)\,\frac{1}{m s}.
\end{equation}

\subsection{Controller Structures}
All controllers are implemented in a one-degree-of-freedom form and evaluated on the same $G(s)$ with the disturbance/noise injections described above. No additional measurement prefiltering is used; the only high-frequency shaping appears as (i) the derivative roll-off in PID and (ii) minimal roll-off factors that guarantee properness in model-based designs.

\subsubsection{PID with Filtered Derivative}
We use the parallel form with a first-order roll-off in the derivative channel,
\begin{equation}
\label{eq:pid}
C_{\mathrm{PID}}(s)
=K_p+\frac{K_i}{s}+\frac{K_d\,s}{1+T_f s},
\end{equation}
where $T_f$ sets the derivative corner frequency. The final coefficients (PSO/DE/WOA) are reported in Table~\ref{tab:final_params}.

\subsubsection{Model-Reference Control (MRC)}
MRC shapes the closed loop to follow a target model $M(s)$. We select a well–damped second-order template augmented with a minimal roll-off to ensure properness~\cite{4}:
\begin{equation}
\label{eq:M}
M(s)=\frac{\omega_n^2}{s^2+2\zeta\omega_n s+\omega_n^2}\cdot\frac{1}{1+\tau_f s},
\qquad \zeta=0.9.
\end{equation}
The associated compensator is
\begin{equation}
\label{eq:Cmrc}
C_{\mathrm{MRC}}(s)=\frac{M(s)}{G(s)\,\big(1-M(s)\big)}.
\end{equation}
The nominal design uses $(\omega_n,\tau_f)=(5.5~\mathrm{rad/s},\,0.10~\mathrm{s})$.

\subsubsection{Energy-Oriented MRC (MRC--R$^{\ast}$)}
To improve actuator friendliness under waves, we reselect $(\omega_n,\tau_f)$ on the \emph{wave} case by a coarse grid that minimizes control energy subject to a soft overshoot cap:
\begin{equation}
\label{eq:mrc_retune}
(\omega_n,\tau_f) \;=\; \arg\min \int u^2\,dt
\quad \text{s.t.}\quad \mathrm{OS}\le 5\%.
\end{equation}
The resulting setting $(\omega_n\approx3.36~\mathrm{rad/s},\,\tau_f\approx0.09~\mathrm{s})$ yields lower $\int u^2$ and $\int(\Delta u)^2$ in the \emph{noise+wave} scenario, with only a mild change in rise/settling times.

\subsubsection{Internal Model Control (IMC)}
IMC is realized through a low-pass design
\begin{equation}
\label{eq:Qimc}
Q(s)=\frac{1}{(\lambda s+1)^n},
\end{equation}
and its equivalent feedback controller
\begin{equation}
\label{eq:Cimc}
C_{\mathrm{IMC}}(s)=\frac{Q(s)}{(1-Q(s))\,G^{-}(s)},
\end{equation}
where $G^{-}$ denotes the invertible, minimum-phase factor of $G$. In our implementation $n=3$ and $\lambda=0.20~\mathrm{s}$ provided a good compromise between tracking and high-frequency attenuation~\cite{5}.

\subsection{Disturbance and Noise Models (for Reproducibility)}
Wave forcing is injected at the actuator as $d(t)=A_w\sin(2\pi f_w t)$ with $(A_w,f_w)=(8~\mathrm{N},\,0.03~\mathrm{Hz})$. Measurement noise is zero-mean white with standard deviation $\sigma=0.12~\mathrm{m/s}$, entering the loop as $y_m=y+n$. These settings are used consistently across all controllers to ensure a fair comparison.

\section{Metaheuristic Tuning and Cost Function}

We tuned the PID parameters $\boldsymbol{\theta}=[K_p,\,K_i,\,K_d,\,T_f]$ with three population--based metaheuristics, all operating on the same bounded, continuous search space
\[
\boldsymbol{\theta}\in [0,\,0,\,0,\,0.005] \times [1000,\,1000,\,1000,\,1000].
\]
Controllers were evaluated in continuous time (MATLAB \texttt{lsim}) on the identified plant, with performance aggregated over four scenarios (nominal, noise, wave, noise+wave).

\paragraph{Particle Swarm Optimization (PSO)}
A swarm of candidate solutions is updated by velocity and position rules that combine individual memory and neighborhood bests. The inertia, cognitive, and social terms balance exploration and exploitation on smooth, low-dimensional landscapes such as PID tuning~\cite{6}.

\paragraph{Differential Evolution (DE)}
DE generates trial solutions by adding the scaled difference of two population members to a third, followed by binomial crossover and greedy selection. Its self-referential mutations are effective for real-valued, nonconvex problems and require few control parameters~\cite{7}.

\paragraph{Whale Optimization Algorithm (WOA)}
WOA alternates between encircling and spiral (``bubble-net'') movements toward elite candidates, with a shrinking search radius to shift from exploration to exploitation. It is lightweight and well suited to continuous parameter vectors~\cite{8}.

\paragraph{Objective}
All algorithms minimized the same scalar cost, designed to prioritize tracking while discouraging energetic and jittery actuation:
\begin{equation}
\label{eq:cost}
\begin{aligned}
J(\boldsymbol{\theta})
&= \sum_{s\in\mathcal{S}} \Big(
  w_T\,\mathrm{ITAE}_s
+ w_I\,\mathrm{IAE}_s
+ w_U \!\int u_s^2\,dt
+ w_D \!\int \dot{u}_s^{\,2}\,dt
\Big) \\
&\quad + \rho\,\big[\max\{0,\ \mathrm{OS}_s-\mathrm{OS}_{\text{ref}}\}\big]^{2}.
\end{aligned}
\end{equation}

Here $\mathrm{IAE}$ and $\mathrm{ITAE}$ are the integral and time-weighted integral of the absolute tracking error, $\int u^2\,dt$ penalizes energy, $\int \dot{u}^{2}\,dt$ penalizes command activity (actuator friendliness), and $\mathrm{OS}_s$ is percent overshoot in scenario $s$; $\mathrm{OS}_{\text{ref}}$ (e.g., $5\%$) sets the soft limit. Scenario weights $(w_T,w_I,w_U,w_D)$ were chosen to make ITAE the primary objective while keeping control effort and smoothness within practical bounds; $\rho$ enforces additional overshoot aversion. 

In each scenario $s$, with error $e_s(t)=r(t)-y(t)$ and horizon $T_s$,
\[
\mathrm{IAE}_s=\int_{0}^{T_s}\!\!\big|e_s(t)\big|\,dt,
\qquad
\mathrm{ITAE}_s=\int_{0}^{T_s}\!\! t\,\big|e_s(t)\big|\,dt .
\]
\subsection{Post Optimization}
The PID controller coefficients obtained after the multi-scenario metaheuristic tuning are summarized in Table~\ref{tab:final_params}. Although the optimizers differ (PSO/DE/WOA), the solutions cluster in a narrow region: PSO and DE effectively yield PI-like settings (either $K_d\!\approx\!0$ or a very large $T_f$), while WOA preserves a small derivative with a shorter roll-off. This consistency indicates a smooth search landscape around the optimum and supports the reproducibility of the reported tunings~\cite{9}.

\begin{table}[htbp]
\caption{Final controller parameters used in the study.}
\label{tab:final_params}
\centering
\begin{tabular}{lcccc}
\hline
\textbf{Controller} & $K_p$ & $K_i$ & $K_d$ & $T_f$ [s] \\
\hline
PID--PSO & 108.842 & 63.386 & 0.067 & 670.8282 \\
PID--DEA & 108.909 & 63.386 & 0.000 & 571.4064 \\
PID--WOA & 122.659 & 0.000 & 0.023 & 3.5256 \\
\hline
\multicolumn{5}{l}{\small PID uses a filtered derivative: $D(s)=\frac{K_d s}{1+T_f s}$.}
\end{tabular}
\end{table}

\subsection{Reference–model/IMC design choices}
The PID gains above are obtained by metaheuristic search against a multi-scenario cost that emphasizes time-weighted tracking while penalizing energy and command activity. In contrast, the model-based controllers (MRC and IMC) are not tuned by stochastic search but by shaping the target closed-loop dynamics. Specifically, the nominal MRC uses a well-damped second-order reference model with a minimal roll-off for properness, whereas the energy-oriented variant (MRC--R$^\ast$) reselects $(\omega_n,\tau_f)$ on the \emph{wave} case via a coarse grid to minimize $\int u^2\,dt$ subject to a soft overshoot cap ($\mathrm{OS}\le5\%$). IMC is realized through a low-pass design $Q(s)=1/(\lambda s+1)^n$ and its equivalent feedback form. All controllers are evaluated on the \emph{same} identified plant and disturbance/noise injections, with measurement noise entering the error channel $e=r-(y+n)$; apart from the derivative roll-off in PID and the minimal properness filter in MRC/IMC, no additional prefiltering is used. This protocol keeps the comparison fair while respecting each method’s design philosophy.

\begin{table}[htbp]
\caption{Reference–model/IMC design choices.}
\label{tab:mrc_imc}
\centering
\begin{tabular}{lccc}
\hline
\textbf{Design} & $\zeta$ & $\omega_n$ [rad/s] & $\tau_f$ [s] \\
\hline
MRC (nominal) & 0.90 & 5.50 & 0.10 \\
MRC--R$^\ast$ (wave–energy) & 0.90 & 3.36 & 0.09 \\
\hline
\end{tabular}

\medskip
\small IMC$^\ast$: low–pass filter $Q(s)=1/(\lambda s+1)^n$ with $n=3$, $\lambda=0.20$ s
(equivalent feedback controller used in simulations).
\end{table}

\section{Results}

\subsection{Evaluation Protocol and Metrics}
We report: rise time, settling time, and percent overshoot (OS) for the step to
$2~\mathrm{m/s}$; RMS error, IAE, and ITAE as aggregate tracking measures;
control energy $\int_0^T u^2 dt$ ($T=50$~s); and actuation activity
$\int_0^T (\Delta u)^2 dt$ as a proxy for actuator stress.
The simulation outcomes for the two representative scenarios are reported next. Tables~\ref{tab:nominal} and~\ref{tab:nw} summarize step–response and actuation metrics, while Figs.~\ref{fig:case1_response}–\ref{fig:case2_error} show the corresponding time traces of the measured speed and absolute tracking error for the nominal and the combined noise+wave cases. Boldface highlights the best (lower-is-better) entry in each column; design choices for MRC and IMC are listed below Table~\ref{tab:nominal}. All controllers were simulated on the same identified plant with identical reference, disturbance, and noise realizations. 
\begin{table*}[t]
\caption{NOMINAL scenario ($T=50$ s). Tracking metrics and actuation costs.}
\label{tab:nominal}
\centering
\begin{tabular}{lcccccccccc}
\hline
\textbf{Controller} & \textbf{Rise [s]} & \textbf{Settle [s]} & \textbf{OS [\%]}
& \textbf{RMS} & \textbf{MAE} & \textbf{IAE} & \textbf{ITAE}
& $\displaystyle\int u^2 dt$ & $\displaystyle\int(\Delta u)^2 dt$ \\
\hline
MRC            & 0.58 & 1.00 & 0.11 & 0.388 & 0.107 & 0.86 & 0.24 & 21987.0 & 357742.0 \\
MRC--R$^\ast$  & 0.89 & 1.51 & 0.14 & 0.465 & 0.157 & 1.25 & 0.52 & \textbf{14162.1} & \textbf{113944.9} \\
IMC$^\ast$     & 0.84 & 1.50 & 0.00 & 0.454 & 0.150 & 1.20 & 0.48 & 15141.0 & 138020.1 \\
PID--PSO       & 0.75 & 4.63 & 20.29 & 0.401 & 0.219 & 1.75 & 2.78 & 19505.7 & \textbf{34176.2} \\
PID--DEA       & 0.75 & 4.63 & 20.29 & 0.401 & 0.219 & 1.75 & 2.78 & 19505.7 & \textbf{34176.2} \\
PID--WOA       & 1.02 & 4.18 & 2.69  & \textbf{0.366} & \textbf{0.144} & \textbf{1.15} & \textbf{1.18} & 16120.4 & 59097.8 \\
\hline
\end{tabular}

\medskip
 MRC (nominal): $n_{\text{roll}}{=}1$, $\tau_f{=}0.10$ s, $\omega_n{=}5.50$ rad/s.
MRC--R$^\ast$ (energy-oriented retune on wave case): $\omega_n{\approx}3.36$ rad/s, $\tau_f{\approx}0.09$ s.
IMC$^\ast$: $Q(s)=1/(\lambda s+1)^n$ with $n=3$, $\lambda=0.20$ s (equivalent feedback used).
\end{table*}

\begin{table*}[t]
\caption{NOISE{+}WAVE scenario ($T=50$ s). Tracking metrics and actuation costs.}
\label{tab:nw}
\centering
\begin{tabular}{lcccccccccc}
\hline
\textbf{Controller} & \textbf{Rise [s]} & \textbf{Settle [s]} & \textbf{OS [\%]}
& \textbf{RMS} & \textbf{MAE} & \textbf{IAE} & \textbf{ITAE}
& $\displaystyle\int u^2 dt$ & $\displaystyle\int(\Delta u)^2 dt$ \\
\hline
MRC            & 0.57 & 46.47 & 4.10 & 0.162 & 0.059 & 2.95 & 53.56 & 24415.6 & 521929.2 \\
MRC--R$^\ast$  & 0.87 & 48.86 & 5.40 & 0.198 & 0.086 & 4.31 & 79.02 & \textbf{16196.3} & \textbf{143991.0} \\
IMC$^\ast$     & 0.82 & 48.82 & 5.20 & 0.193 & 0.082 & 4.12 & 75.74 & 17215.0 & 169479.8 \\
PID--PSO       & 0.74 & 49.05 & 21.39 & 0.162 & \textbf{0.047} & \textbf{2.37} & \textbf{20.95} & 30405.5 & $6.824\times10^{8}$ \\
PID--DEA       & 0.74 & 49.05 & 21.39 & 0.162 & \textbf{0.047} & \textbf{2.37} & \textbf{20.95} & 30405.5 & $6.824\times10^{8}$ \\
PID--WOA       & 1.00 & 46.35 & 4.68  & \textbf{0.154} & 0.064 & 3.18 & 52.03 & 29091.9 & $8.686\times10^{8}$ \\
\hline
\end{tabular}

\medskip

\end{table*}

\subsection{Discussion}
In Nominal state, PID--WOA delivers the best tracking aggregates (lowest RMS/IAE/ITAE) with small overshoot,
while MRC--R$^\ast$ and IMC$^\ast$ achieve substantially lower energy and command activity than nominal MRC, at a modest
cost in tracking indices. PID--PSO/DEA settle slower and overshoot more due to aggressive gains, but yield the
smoothest commands (smallest $\int(\Delta u)^2 dt$) when no noise is present.

Under simultaneous sensor noise and wave forcing, model-based designs (MRC--R$^\ast$, IMC$^\ast$)
provide the best \emph{actuator economy and smoothness}---$\int u^2 dt$ reduced by $\sim$34\% vs.\ nominal MRC, and activity
cut by $\sim$3.6$\times$ (MRC--R$^\ast$). PID--PSO/DEA retain strong ITAE/MAE but incur extremely large activity,
highlighting sensitivity to $n(t)$ entering the error. PID--WOA keeps overshoot within 5\% with competitive RMS, at the
expense of slow dynamics comparable to MRC/IMC.
 If energy and actuator friendliness are primary mission drivers under noise+wave, \textbf{MRC--R$^\ast$}
(and closely \textbf{IMC}$^\ast$) are preferred. If the priority is minimal ITAE and MAE and a $\sim$20\% overshoot is acceptable,
\textbf{PID--PSO/DEA} are viable alternatives; a conservative, small-overshoot option with good RMS is \textbf{PID--WOA}.
\begin{figure*}[ht]
    \centering
    
    \begin{minipage}[b]{0.48\textwidth}
        \centering
        \input{case1_response.tex}
        \captionof{figure}{Case-1: System Response}
        \label{fig:case1_response}
    \end{minipage}
    \hfill
    \begin{minipage}[b]{0.48\textwidth}
        \centering
        \input{case1_error.tex}
        \captionof{figure}{Case-1: Absolute Error}
        \label{fig:case1_error}
    \end{minipage}
    
    \vspace{1cm} 
    
    \begin{minipage}[b]{0.48\textwidth}
        \centering
        \input{case2_response.tex}
        \captionof{figure}{Case-2: System Response (Disturbed)}
        \label{fig:case2_response}
    \end{minipage}
    \hfill
    \begin{minipage}[b]{0.48\textwidth}
        \centering
        \input{case2_error.tex}
        \captionof{figure}{Case-2: Absolute Error (Disturbed)}
        \label{fig:case2_error}
    \end{minipage}
\end{figure*}

\end{document}

%% file: case1_response.tex
%
\definecolor{mycolor1}{rgb}{0.06600,0.44300,0.74500}%
\definecolor{mycolor2}{rgb}{0.86600,0.32900,0.00000}%
\definecolor{mycolor3}{rgb}{0.92900,0.69400,0.12500}%
\definecolor{mycolor4}{rgb}{0.52100,0.08600,0.81900}%
\definecolor{mycolor5}{rgb}{0.23100,0.66600,0.19600}%
\definecolor{mycolor6}{rgb}{0.18400,0.74500,0.93700}%
\definecolor{mycolor7}{rgb}{0.12941,0.12941,0.12941}%
\begin{tikzpicture}

\begin{axis}[%
width=0.9\linewidth,
height=0.23\textheight,
at={(0\linewidth,0\textheight)},
scale only axis,
xmin=0,
xmax=50,
xlabel style={font=\color{mycolor7}},
xlabel={Time [s]},
ymin=1.6,
ymax=2.1,
ylabel style={font=\color{mycolor7}},
ylabel={y(t) [m/s]},
axis background/.style={fill=white},
xmajorgrids,
ymajorgrids,
legend style={legend cell align=left, align=left},
xmin=0,xmax=50,
tick label style={font=\footnotesize},
label style={font=\small},
legend style={font=\tiny},
title style={font=\small}
]
\addplot [color=mycolor1, line width=1.0pt]
  table[row sep=crcr]{%
0	1.8\\
0.350000000000001	1.8346728786757\\
0.549999999999997	1.85379372983846\\
0.75	1.87209441752528\\
0.950000000000003	1.88939242911989\\
1.1	1.90162722773061\\
1.25	1.91318454443548\\
1.4	1.92403812201497\\
1.55	1.93417270567705\\
1.7	1.94358266972285\\
1.85	1.95227073500686\\
2	1.96024677793072\\
2.15	1.96752673065146\\
2.3	1.9741315712773\\
2.45	1.98008640205602\\
2.6	1.98541961292009\\
2.8	1.9916173237182\\
3	1.99684327937268\\
3.2	2.00117905637595\\
3.4	2.00470755325454\\
3.6	2.00751125073088\\
3.85	2.01011921400938\\
4.1	2.01187122969427\\
4.4	2.01304155649833\\
4.7	2.01340343148196\\
5.05	2.0130631711998\\
5.5	2.01181773483485\\
6.25	2.00878617709768\\
7.4	2.0041936047167\\
8.15	2.00202241040131\\
8.95	2.00053571776726\\
9.85	1.99969349429411\\
11.05	1.99942118410823\\
14.2	1.99988703543507\\
18.75	2.00001694263474\\
50	1.99999999999725\\
};
\addlegendentry{MRC}

\addplot [color=mycolor2, line width=1.0pt]
  table[row sep=crcr]{%
0	1.75\\
0.399999999999999	1.78968510054337\\
0.649999999999999	1.81371586871368\\
0.850000000000001	1.83224380185774\\
1.05	1.85001445388988\\
1.25	1.86692981744608\\
1.45	1.88291648215488\\
1.65	1.89792286522813\\
1.85	1.91191659361384\\
2.05	1.92488204623381\\
2.25	1.93681806167721\\
2.45	1.94773581397205\\
2.65	1.95765685668282\\
2.85	1.96661133355704\\
3.05	1.97463635223809\\
3.25	1.98177451615011\\
3.45	1.98807260851526\\
3.65	1.99358042155922\\
3.85	1.99834972327184\\
4.05	2.0024333535932\\
4.3	2.00665457707048\\
4.55	2.00999019630682\\
4.8	2.0125403802373\\
5.1	2.01469897086329\\
5.4	2.01602151782553\\
5.75	2.01670001587344\\
6.15	2.01658656102408\\
6.6	2.01564270945707\\
7.2	2.01355625490103\\
9.8	2.00353187714678\\
10.65	2.00154326911257\\
11.55	2.0002272488344\\
12.65	1.9994595920236\\
14.15	1.9992930285616\\
27.4	2.00000015456711\\
50	1.99999999978972\\
};
\addlegendentry{MRC-R*}

\addplot [color=mycolor3, line width=1.0pt]
  table[row sep=crcr]{%
0	1.7\\
0.200000000000003	1.72141394804635\\
0.5	1.75451067290496\\
0.950000000000003	1.80422277940338\\
1.2	1.83093357157464\\
1.4	1.85149430635428\\
1.6	1.87116852597838\\
1.8	1.88983363634084\\
1.95	1.90311249374718\\
2.1	1.91573992758653\\
2.25	1.92769326653777\\
2.4	1.93895677555867\\
2.55	1.94952107367017\\
2.7	1.95938256569219\\
2.85	1.96854289048282\\
3	1.97700838786351\\
3.15	1.98478958606611\\
3.3	1.99190071121187\\
3.45	1.99835922002947\\
3.6	2.00418535673738\\
3.8	2.01100922581914\\
4	2.01680820402575\\
4.2	2.0216459135352\\
4.4	2.02558931335928\\
4.6	2.0287074520119\\
4.8	2.03107035781053\\
5	2.03274806227763\\
5.25	2.03398688791986\\
5.5	2.03439457250515\\
5.8	2.0339633987509\\
6.1	2.03272002325349\\
6.45	2.03049074755835\\
6.9	2.02679225395045\\
7.7	2.01921225635665\\
8.5	2.01187504757322\\
9.1	2.00718626086677\\
9.65	2.00370181122099\\
10.2	2.00103121093222\\
10.75	1.99913053870593\\
11.35	1.99782606596901\\
12.05	1.997118111174\\
12.9	1.99708517696027\\
14.2	1.99792892684685\\
16.95	1.99982135820121\\
18.95	2.00024641471359\\
23.05	2.00005489301248\\
31.6	2.00000039041926\\
50	2.0000000065786\\
};
\addlegendentry{IMC*}

\addplot [color=mycolor4, line width=1.0pt]
  table[row sep=crcr]{%
0	1.65\\
0.100000000000001	1.66600297911013\\
0.25	1.69075287575055\\
0.5	1.7330019116712\\
0.75	1.77511946719913\\
0.899999999999999	1.79977486967042\\
1.05	1.82368468080733\\
1.15	1.83911573046122\\
1.25	1.85408368731412\\
1.35	1.86854792980067\\
1.45	1.88247404096694\\
1.55	1.89583345949925\\
1.65	1.90860312930405\\
1.75	1.92076514964772\\
1.85	1.93230642768913\\
1.95	1.9432183350674\\
2.05	1.95349637004215\\
2.15	1.96313982652502\\
2.25	1.97215147118892\\
2.35	1.98053722969577\\
2.45	1.98830588294508\\
2.55	1.99546877411405\\
2.65	2.00203952713565\\
2.75	2.00803377714371\\
2.85	2.01346891330461\\
2.95	2.01836383435208\\
3.05	2.02273871704611\\
3.15	2.02661479768881\\
3.25	2.03001416674793\\
3.35	2.03295957656423\\
3.5	2.03657743480784\\
3.65	2.03930582639187\\
3.8	2.04122480905448\\
3.95	2.04241390311607\\
4.1	2.04295121888705\\
4.25	2.04291271094599\\
4.45	2.04209164643316\\
4.65	2.04054074020458\\
4.9	2.03780624172118\\
5.2	2.03370681407606\\
5.6	2.0274918665988\\
6.35	2.01577015337784\\
6.7	2.01095842135021\\
7.05	2.00680283977173\\
7.4	2.0033703355443\\
7.75	2.0006672554812\\
8.1	1.99865536271498\\
8.45	1.99726613090792\\
8.85	1.99632934661038\\
9.3	1.99593604855775\\
9.85	1.99612664694924\\
10.65	1.9971144440523\\
12.5	1.99956244324437\\
13.6	2.00023181076902\\
15	2.00037338834944\\
24.8	2.00000336923007\\
50	1.99999999999086\\
};
\addlegendentry{PID-PSO}

\addplot [color=mycolor5, line width=1.0pt]
  table[row sep=crcr]{%
0	1.72\\
0.25	1.75827359229476\\
0.399999999999999	1.78071070656799\\
0.549999999999997	1.80248368652144\\
0.700000000000003	1.82342063671916\\
0.799999999999997	1.83684595351822\\
0.899999999999999	1.84980706904278\\
1	1.86227811882286\\
1.1	1.87423873881512\\
1.2	1.8856736090555\\
1.3	1.896572015964\\
1.4	1.90692743386522\\
1.5	1.91673712611918\\
1.6	1.92600176610386\\
1.7	1.93472507815174\\
1.8	1.94291349841817\\
1.9	1.95057585554697\\
2	1.95772307089995\\
2.1	1.96436787802887\\
2.2	1.97052456099172\\
2.3	1.9762087110491\\
2.4	1.98143700121906\\
2.55	1.98846327383828\\
2.7	1.99456541379521\\
2.85	1.99980749992405\\
3	2.00425464465029\\
3.15	2.00797202062018\\
3.3	2.01102403463115\\
3.45	2.0134736368122\\
3.6	2.01538175314096\\
3.8	2.01718421192236\\
4	2.01825817909766\\
4.2	2.01872423532123\\
4.45	2.01861839007792\\
4.75	2.01772997146129\\
5.1	2.01599120985311\\
5.65	2.01251225214732\\
6.55	2.00683518317881\\
7.1	2.00407442719458\\
7.6	2.00217431683731\\
8.15	2.00072130521924\\
8.75	1.99977255727653\\
9.5	1.99926113016007\\
10.55	1.99925022603243\\
15.95	2.0000344758212\\
50	2.00000000000023\\
};
\addlegendentry{PID-DEA}

\addplot [color=mycolor6, line width=1.0pt]
  table[row sep=crcr]{%
0	1.7\\
0.200000000000003	1.72554862114956\\
0.700000000000003	1.78997228012346\\
0.899999999999999	1.81489585436952\\
1.05	1.83297706614105\\
1.2	1.85042210330838\\
1.35	1.86714915237808\\
1.5	1.88309255226494\\
1.65	1.89820147349482\\
1.8	1.91243862018225\\
1.95	1.92577896377774\\
2.1	1.93820851622905\\
2.25	1.94972314894034\\
2.4	1.9603274627451\\
2.55	1.9700337130317\\
2.7	1.97886079317369\\
2.85	1.98683327851865\\
3	1.99398053237709\\
3.15	2.00033587472401\\
3.3	2.00593581367612\\
3.45	2.01081933923533\\
3.6	2.01502727828812\\
3.75	2.01860170941869\\
3.9	2.02158543572558\\
4.05	2.02402151352316\\
4.2	2.02595283455723\\
4.4	2.0278158505213\\
4.6	2.02895466198257\\
4.8	2.0294634101319\\
5.05	2.02934910407428\\
5.3	2.02855252685902\\
5.6	2.0269038250814\\
5.95	2.02430886875739\\
6.5	2.01945092939359\\
7.45	2.01095688876758\\
7.95	2.00715504232556\\
8.4	2.00431874449394\\
8.85	2.00206692154616\\
9.35	2.00022094758322\\
9.9	1.99889567470533\\
10.5	1.99813265291119\\
11.2	1.99789225190417\\
12.2	1.99824422674792\\
15.8	2.00001811010518\\
18.15	2.00013783755073\\
31.75	2.0000004783186\\
50	1.99999999977452\\
};
\addlegendentry{PID-WOA}

\addplot [color=black, dashed, line width=1.2pt, forget plot]
  table[row sep=crcr]{%
0	2\\
50	2\\
};
\end{axis}
\end{tikzpicture}%

%% file: case1_error.tex
%
\definecolor{mycolor1}{rgb}{0.06600,0.44300,0.74500}%
\definecolor{mycolor2}{rgb}{0.86600,0.32900,0.00000}%
\definecolor{mycolor3}{rgb}{0.92900,0.69400,0.12500}%
\definecolor{mycolor4}{rgb}{0.52100,0.08600,0.81900}%
\definecolor{mycolor5}{rgb}{0.23100,0.66600,0.19600}%
\definecolor{mycolor6}{rgb}{0.18400,0.74500,0.93700}%
\definecolor{mycolor7}{rgb}{0.12941,0.12941,0.12941}%
\begin{tikzpicture}

\begin{axis}[%
width=1\linewidth,
height=0.23\textheight,
at={(0\linewidth,0\textheight)},
scale only axis,
xmin=0,
xmax=50,
xlabel style={font=\color{mycolor7}},
xlabel={Time [s]},
ymin=0,
ymax=0.4,
ylabel style={font=\color{mycolor7}},
ylabel={$|$e(t)$|$ [m/s]},
axis background/.style={fill=white},
xmajorgrids,
ymajorgrids,
legend style={legend cell align=left, align=left},
xmin=0,xmax=50,
tick label style={font=\footnotesize},
label style={font=\small},
legend style={font=\tiny},
title style={font=\small}
]
\addplot [color=mycolor1, line width=1.0pt]
  table[row sep=crcr]{%
0	0.200000000000003\\
0.299999999999997	0.170208630727593\\
0.5	0.150917950551175\\
0.649999999999999	0.136940714306697\\
0.799999999999997	0.123481129580327\\
0.950000000000003	0.110607570880113\\
1.1	0.0983727722693928\\
1.25	0.0868154555645191\\
1.4	0.07596187798503\\
1.55	0.0658272943229505\\
1.7	0.0564173302771493\\
1.85	0.0477292649931371\\
2	0.0397532220692796\\
2.15	0.032473269348543\\
2.3	0.0258684287226956\\
2.45	0.019913597943976\\
2.6	0.0145803870799099\\
2.75	0.00983787277053239\\
2.9	0.00565327386532033\\
3.05	0.00199255233922457\\
3.1	0.000882732013458565\\
3.15	0.00017403790792514\\
3.3	0.00303903189306709\\
3.5	0.00619490399893863\\
3.7	0.00866660589487367\\
3.9	0.0105334268199258\\
4.1	0.0118712296942718\\
4.35	0.0129083155522522\\
4.6	0.0133600126987474\\
4.9	0.0132932034691322\\
5.25	0.0125973777929858\\
5.7	0.0110805104597702\\
6.6	0.00728427633178086\\
7.4	0.00419360471669705\\
8.05	0.002266944538448\\
8.7	0.000915116412130601\\
9.4	2.25356654226516e-05\\
9.8	0.000277775970531025\\
10.8	0.000573328349581459\\
12.5	0.000385878252672001\\
16	9.76196474056223e-06\\
25.8	4.72061586265227e-07\\
50	2.74980038739159e-12\\
};
\addlegendentry{MRC}

\addplot [color=mycolor2, line width=1.0pt]
  table[row sep=crcr]{%
0	0.25\\
0.350000000000001	0.215213107782773\\
0.549999999999997	0.195794011992859\\
0.75	0.176932669506584\\
0.950000000000003	0.158769476592759\\
1.1	0.145673255787301\\
1.25	0.133070182553915\\
1.4	0.120989949700771\\
1.55	0.109455347159937\\
1.7	0.0984828880297854\\
1.85	0.0880834063861613\\
2	0.0782626260011199\\
2.15	0.0690216994026045\\
2.3	0.0603577169738045\\
2.45	0.052264186027962\\
2.6	0.0447314800053249\\
2.75	0.0377472581250444\\
2.9	0.0312968559872573\\
3.05	0.0253636477619068\\
3.2	0.0199293807212371\\
3.35	0.014974482975056\\
3.5	0.0104783453522543\\
3.65	0.00641957844077723\\
3.85	0.00165027672816365\\
3.9	0.000567163405712279\\
3.95	0.000473925182156165\\
4.15	0.00423465257955513\\
4.35	0.00738929688172618\\
4.55	0.0099901963068163\\
4.75	0.0120884935312162\\
5	0.0140798374980307\\
5.25	0.0154554126966815\\
5.5	0.0163019456229136\\
5.8	0.0167324508435911\\
6.15	0.0165865610240843\\
6.55	0.0157805068672872\\
7.05	0.0141375344721055\\
7.85	0.0108206857518951\\
9	0.00612569451755007\\
9.75	0.00367356435431532\\
10.45	0.00194135431902254\\
11.2	0.0006525295943689\\
11.8	1.6963235928813e-05\\
12.75	0.00057610862618418\\
14	0.000716984230471951\\
16.75	0.000281381275193837\\
20.1	1.49025527420577e-05\\
30.7	1.11364578003759e-06\\
50	2.10278017220844e-10\\
};
\addlegendentry{MRC-R*}

\addplot [color=mycolor3, line width=1.0pt]
  table[row sep=crcr]{%
0	0.299999999999997\\
0.149999999999999	0.284008059875049\\
0.350000000000001	0.262126635528105\\
1	0.190360304534899\\
1.2	0.16906642842536\\
1.35	0.153568212325176\\
1.5	0.13854941548226\\
1.65	0.124066443806832\\
1.8	0.110166363659161\\
1.95	0.0968875062528198\\
2.1	0.0842600724134712\\
2.25	0.0723067334622343\\
2.4	0.0610432244413275\\
2.55	0.0504789263298306\\
2.7	0.0406174343078121\\
2.85	0.0314571095171843\\
3	0.0229916121364866\\
3.15	0.0152104139338931\\
3.3	0.00809928878813082\\
3.45	0.00164077997052914\\
3.5	0.000370457321039908\\
3.65	0.00599072309858428\\
3.8	0.0110092258191443\\
3.95	0.015451130857123\\
4.1	0.0193430953377245\\
4.25	0.0227129345132724\\
4.4	0.0255893133592835\\
4.55	0.0280014630158334\\
4.75	0.0305465962281346\\
4.95	0.0323890684827575\\
5.15	0.0335983497727881\\
5.35	0.0342425073466757\\
5.55	0.0343875641339153\\
5.8	0.0339633987508989\\
6.05	0.0329755108123422\\
6.35	0.0311986041905783\\
6.7	0.0285240653967165\\
7.15	0.0244932786998149\\
8.6	0.0110349608901998\\
9.1	0.00718626086677432\\
9.55	0.00427456746633226\\
10	0.00190964447376984\\
10.45	7.65303367202819e-05\\
10.55	0.00026232705652518\\
11.05	0.00161338410709533\\
11.6	0.00251537729680962\\
12.2	0.00294279786538709\\
12.95	0.00289693224475229\\
14.05	0.00219352415348339\\
16.8	0.000242949558838745\\
17.75	8.00417646331653e-05\\
19.65	0.000258260349752959\\
50	6.57860255159903e-09\\
};
\addlegendentry{IMC*}

\addplot [color=mycolor4, line width=1.0pt]
  table[row sep=crcr]{%
0	0.350000000000001\\
0.100000000000001	0.333997020889875\\
0.25	0.309247124249445\\
0.450000000000003	0.275484532184223\\
0.700000000000003	0.233226752450314\\
0.850000000000001	0.208371121601608\\
1	0.184190027910866\\
1.1	0.16854469845881\\
1.2	0.153339716173562\\
1.3	0.138618934743896\\
1.4	0.124419829981989\\
1.5	0.110773847260532\\
1.6	0.0977067514706178\\
1.7	0.0852389774019144\\
1.8	0.0733859786248203\\
1.9	0.0621585731276397\\
2	0.0515632841296849\\
2.1	0.041602674653042\\
2.2	0.0322756745912613\\
2.3	0.0235778991621345\\
2.4	0.015501957773921\\
2.5	0.00803775246940575\\
2.6	0.00117276524012055\\
2.65	0.00203952713565059\\
2.75	0.00803377714370868\\
2.85	0.0134689133046137\\
2.95	0.0183638343520798\\
3.05	0.02273871704611\\
3.15	0.0266147976888149\\
3.25	0.0300141667479323\\
3.35	0.032959576564231\\
3.45	0.0354742620505348\\
3.55	0.0375817742283004\\
3.7	0.0400314746810864\\
3.85	0.041698378475381\\
4	0.0426616264761961\\
4.15	0.0429986872113872\\
4.3	0.0427846518676347\\
4.45	0.0420916464331569\\
4.65	0.0405407402045839\\
4.85	0.0384120941876276\\
5.1	0.0351491618837372\\
5.45	0.0298793072419272\\
6.45	0.0143352161165922\\
6.8	0.00969992490477978\\
7.1	0.00626748528650722\\
7.4	0.00337033554429667\\
7.7	0.00100981037078185\\
7.85	2.44418997112916e-05\\
8.4	0.00257037618590061\\
8.75	0.00349442707197767\\
9.15	0.00399886856970966\\
9.65	0.00400899955289447\\
10.3	0.00337906849003389\\
13	5.43851595864453e-05\\
13.3	0.000110687690764166\\
14.45	0.000382342200303754\\
16.5	0.000185535341984178\\
19.55	3.5443900102905e-05\\
50	9.13757958187489e-12\\
};
\addlegendentry{PID-PSO}

\addplot [color=mycolor5, line width=1.0pt]
  table[row sep=crcr]{%
0	0.280000000000001\\
0.25	0.241726407705244\\
0.399999999999999	0.219289293432006\\
0.549999999999997	0.197516313478559\\
0.649999999999999	0.183456394130744\\
0.75	0.16981052396109\\
0.850000000000001	0.156613712642255\\
0.950000000000003	0.143894751617445\\
1.05	0.131676696961051\\
1.15	0.11997733478384\\
1.25	0.108809628335784\\
1.35	0.0981821461528938\\
1.45	0.0880994707709988\\
1.55	0.0785625876901577\\
1.65	0.0695692544198536\\
1.75	0.0611143495664166\\
1.85	0.0531902020427637\\
1.95	0.0457869005859024\\
2.05	0.0388925838609637\\
2.15	0.03249371151297\\
2.25	0.0265753165986382\\
2.35	0.0211212398924516\\
2.45	0.0161143466132643\\
2.55	0.0115367261617223\\
2.65	0.00736987549431234\\
2.75	0.00359486678862453\\
2.85	0.000192500075954172\\
2.9	0.00137498262264302\\
3.05	0.00557165186398123\\
3.2	0.0090601431579671\\
3.35	0.0119044481698438\\
3.5	0.0141668996131301\\
3.65	0.015907642274243\\
3.8	0.0171842119223555\\
4	0.0182581790976641\\
4.2	0.0187242353212298\\
4.45	0.0186183900779184\\
4.7	0.0179244253269246\\
5	0.0165448287687866\\
5.45	0.0138304423845454\\
6.6	0.00655646492342754\\
7.1	0.00407442719458118\\
7.6	0.00217431683731206\\
8.1	0.000828247552000505\\
8.55	2.51207538042308e-05\\
9	0.000466844844133618\\
9.75	0.000794705928321093\\
10.9	0.000668764279893708\\
14.45	9.24413603797802e-06\\
20.1	2.62876213241725e-07\\
50	2.27373675443232e-13\\
};
\addlegendentry{PID-DEA}

\addplot [color=mycolor6, line width=1.0pt]
  table[row sep=crcr]{%
0	0.299999999999997\\
0.200000000000003	0.274451378850436\\
0.700000000000003	0.210027719876535\\
0.850000000000001	0.191255954907668\\
1	0.172983849424568\\
1.15	0.155316827672642\\
1.3	0.138342383254084\\
1.45	0.122131413796566\\
1.6	0.106739547096637\\
1.7	0.0969544706752359\\
1.8	0.0875613798177497\\
1.9	0.0785670951708468\\
2	0.0699762573828835\\
2.1	0.0617914837709534\\
2.2	0.0540135206376178\\
2.3	0.0466413909649788\\
2.4	0.0396725372549014\\
2.5	0.0331029593223064\\
2.6	0.026927346884861\\
2.7	0.0211392068263052\\
2.8	0.0157309850421257\\
2.95	0.00831217726504718\\
3.1	0.00169677855916461\\
3.15	0.000335874724008534\\
3.3	0.00593581367612472\\
3.45	0.0108193392353328\\
3.6	0.015027278288116\\
3.75	0.0186017094186894\\
3.9	0.0215854357255836\\
4.05	0.0240215135231594\\
4.2	0.0259528345572306\\
4.35	0.0274217591623724\\
4.55	0.0287326310172347\\
4.75	0.0293903120761598\\
4.95	0.0294855046833149\\
5.2	0.0289432002252852\\
5.45	0.0278094296355391\\
5.75	0.0258631084150025\\
6.15	0.0226138724495684\\
6.95	0.015304408624992\\
7.55	0.010146512458924\\
8	0.00681134412847939\\
8.45	0.00403973898549737\\
8.9	0.00185216299804125\\
9.35	0.000220947583216002\\
9.45	7.1473515795617e-05\\
9.95	0.00119265330094009\\
10.5	0.00186734708881175\\
11.15	0.00210801089023249\\
12.05	0.00183873095879505\\
15.9	3.61900746526089e-05\\
18.15	0.00013783755073149\\
30.55	7.28859390619618e-07\\
50	2.25476526338753e-10\\
};
\addlegendentry{PID-WOA}

\end{axis}
\end{tikzpicture}%

%% file: conference_101719.bbl
\begin{thebibliography}{00}
\bibitem{1} L.-Y. Hao, Z.-J. Wu, C. Shen, Y. Cao, and R.-Z. Wang, 
“Tube-based model predictive control for constrained unmanned marine vehicles with thruster faults,” 
IEEE Trans. Ind. Informat., vol. 20, no. 3, pp. 4606–4615, 2023.
\bibitem{2} H. Ersoy, B. Akgül, E. Akpınar, A. Kartci, and U. E. Ayten, 
“A robust speed controller design for PMSM using  PI$^{\lambda}$D$^{\mu}$ controllers,” 
in Proc. 47th Int. Conf. Telecommun. Signal Process. (TSP), 2024, pp. 289–293.

\bibitem{1.matlab}MathWorks Student Competitions Team, MATLAB and Simulink
Robotics Arena: From Data to Model (2024).
\bibitem{4} S. Baek and J. Woo, 
“Model reference adaptive control-based autonomous berthing of an unmanned surface vehicle under environmental disturbance,” 
Machines, vol. 10, no. 4, p. 244, 2022.
\bibitem{5}N. Abe and K. Yamanaka, “Smith predictor control and internal model control — a tutorial,” in *Proc. SICE 2003 Annual Conf.* (IEEE Cat. No. 03TH8734), vol. 2, 2003, pp. 1383–1387.
\bibitem{6} J. Kennedy and R. C. Eberhart, 
“Particle swarm optimization,” 
in Proc. IEEE Int. Conf. Neural Networks (ICNN’95), 
Perth, WA, Australia, vol. 4, 1995, pp. 1942–1948.

\bibitem{7} R. Storn and K. Price, 
“Differential evolution—a simple and efficient heuristic for global optimization over continuous spaces,” 
J. Global Optim., vol. 11, no. 4, pp. 341–359, 1997.
\bibitem{8} S. Mirjalili and A. Lewis, 
“The whale optimization algorithm,” 
Adv. Eng. Softw., vol. 95, pp. 51–67, 2016.
\bibitem{9}H. Ersoy, B. Akgül, E. Akpinar, A. Kartci and U. E. Ayten, "PI$^{\lambda}$D$^{\mu}$ Approach for High-Efficiency BLDC Motor Control," 2024 15th National Conference on Electrical and Electronics Engineering (ELECO), Bursa, Turkiye, 2024, pp. 1-4, doi: 10.1109/ELECO64362.2024.10847114.

\end{thebibliography}
